# Pulse Laser Deposited $GaFeO_3$ Thin Film Grown on FTO: Correlated Ferroic Orders and Strong Magnetodielectric Coupling at Room Temperature

Ishita Jana*, Sourav Sarkar, Soham Saha and Kalyan Mandal
S. N. Bose National Centre for Basic Sciences, JD Block, Sector III, Salt Lake City, Kolkata 700106, India
*Corresponding author's email address: ishita.jana@bose.res.in

## ABSTRACT:

Room-temperature single-phase multiferroics exhibiting coupled ferroelectric and magnetic order are promising candidates for next-generation magnetoelectric memory and spintronic technologies. Here, we report the growth of ~100.6 nm thick polycrystalline orthorhombic $GaFeO_3$ thin films on fluorine-doped tin oxide-coated glass (FTO) substrates by pulsed laser deposition. Core-level X-ray photoelectron spectroscopy reveals a reduced $Fe^{2+}$ concentration and suppressed oxygen-vacancy formation, contributing to reduced leakage current and improved ferroelectric characteristics. The films exhibit robust ferrimagnetism and ferroelectricity at room temperature, confirming the coexistence of ferroic orders within a single phase. Piezoresponse Force Microscopy further demonstrates reversible local polarization switching, accompanied by an asymmetric butterfly loop with coercive voltages of −0.31V and +1.07 V. The films also exhibit a high relative dielectric constant of ~1700 with a relatively low dielectric loss tangent of ~2.83 at 1 MHz. Moreover, a significant magnetodielectric coefficient of ~2.3% is observed under a relatively small magnetic field of only 0.1 kOe, highlighting strong magnetoelectric coupling. These results establish $GaFeO_3$ thin films as a promising platform for room-temperature multifunctional devices, mainly low-field magnetoelectric memory and spintronic architectures.

## INTRODUCTION

Multiferroic materials, in which electric and magnetic orders coexist within a single phase, have attracted considerable interest because the coupling between these order parameters adds additional functionalities beyond those of conventional ferroelectric and magnetic materials [1–4]. However, only a limited number of single-phase multiferroics exhibit significant magnetoelectric (ME) coupling at room temperature (RT), which remains an important challenge for practical application. $GaFeO_3$ (GFO) is an interesting multiferroic because its non-centrosymmetric orthorhombic structure ($Pc2_1n$) permits polar order, while the Fe sublattice gives rise to ferrimagnetic ordering. The four crystallographically inequivalent cation sites Ga1, Ga2, Fe1, and Fe2, consisting of one tetrahedral and three octahedral sites, allow partial Ga/Fe site disorder, which is strongly associated with the magnetic properties of GFO[5,6].

Despite its multifunctional properties, achieving robust ferroelectric behaviour in bulk GFO remains challenging. Polarization hysteresis measurements of bulk polycrystalline and single-crystalline GFO are often affected by significant leakage currents, leading to weak or unclear ferroelectric responses[7,8]. These limitations motivate the use of thin-film GFO, where microstructure, interfaces, crystallographic orientation, and substrate effects can be controlled more effectively[9]. Previous studies have reported room-temperature ferrimagnetism and nanoscale polarization switching in both polycrystalline and epitaxial GFO thin films, demonstrating the potential of thin-film GFO for studying its coupled ferroic properties.

However, the fabrication of GFO thin films has often involved single-crystalline substrates and complex growth procedures, limiting the range of substrate platforms explored for device integration. One such approach involves the demonstration of room-temperature ferroelectricity in epitaxial GFO films deposited by pulsed laser deposition (PLD) on yttria-stabilized zirconia (YSZ) and $SrTiO_3$ (STO) substrates[10,11]. Furthermore, metal–insulator–metal capacitor structures in these studies required additional deposition of conducting electrodes. Although such approaches are useful for investigating the intrinsic properties of GFO, the use of expensive substrates and multiple deposition steps makes device fabrication more complex and costly. Therefore, developing GFO thin films directly on commercially available conductive substrates is of practical interest, especially when the substrate can simultaneously serve as an electrode.

Among conductive substrates, FTO offers a practical route for electrical measurements while eliminating the need for a separate bottom-electrode layer. However, the structural, chemical, electrical, and magnetic properties of GFO grown directly on FTO remain insufficiently explored. This is particularly relevant because the substrate and thin-film environment can modify the structural and electronic properties of GFO. Previous studies have shown that substrate-induced strain and structural changes can influence the properties of GFO thin films. Therefore, investigating GFO directly on FTO provides a promising approach to investigate its coupled ferroic properties in a simple and device-relevant configuration.

In the present work, we investigate polycrystalline GFO thin films deposited directly on FTO substrates by PLD. The structural and morphological characteristics are correlated with the surface chemical states obtained from X-ray photoelectron spectroscopy. Room-temperature local polarization switching is examined using Piezoresponse Force Microscopy (PFM), along with the dielectric and magnetic properties of the films. Finally, the field-dependent dielectric response is investigated to assess the magnetodielectric behaviour. The combined observation of room-temperature ferrimagnetic ordering, local polarization switching, and a measurable low-field magnetodielectric response provides a platform for exploring coupled ferroic behaviour in GFO thin films fabricated on a commercially available conductive substrate.

## EXPERIMENTAL DETAILS

Using the pulsed laser deposition (PLD) approach in oxygen atmospheres with the KrF excimer laser (248 nm), GFO thin films are deposited on fluorine-doped tin oxide-coated glass (FTO) substrates at 600 ºC. With a fluence of 2 $J.cm^{-2}$ and an oxygen pressure of 0.04 mbar, the thin films are formed at a repetition rate of 5 Hz with a total of 7000 shots. The polycrystalline GFO powder, which is produced via the sol-gel process, is used to prepare the PLD targets. In a nutshell, the initial reagents are $Ga(NO_3)_3.H_2O$, and $Fe(NO_3)_3.9H2O$, which are then successively dissolved in double-distilled water at their stoichiometric ratios. The solution is stirred to achieve homogeneity and then heated at a temperature of 80 ºC until completely dried. The dried powder is grinded, and calcined several times at a temperature of 800 ºC for 6 hours each. To be used as the PLD target, the powders are transformed into pellets, and the final calcination is performed at 1300 ºC for 4 hours.

The phase and structure of the as-prepared samples are investigated at room temperature by X-ray diffraction (XRD) using a Rigaku SmartLab with Cu–Kα (λ = 1.54 Å) in the 2θ range of 20◦–70◦. Field emission scanning electron microscope (FESEM) and energy-dispersive X-ray (EDAX) spectra are studied using FEI QUANTA FEG-250 with the attachment of AMETEK EDAX QUANTA 200. X-ray photoelectron spectroscopy (XPS) measurements are performed using a ULVAC-PHI 5000 Versa Probe III system under ultra-high vacuum conditions (5 × 10−10 mbar). An Al Kα X-ray source (100 µm, 250 W, 15 kV) was employed, with the analyzer pass energy set to 26 eV and a step size of 0.1 eV. The field-dependent magnetization is studied using the vibrating sample magnetometer (Lakeshore VSM and Quantum Design PPMS). The RT ferroelectric hysteresis measurement is performed at a frequency of 100 Hz using a precision impedance analyzer II (Radiant Technologies). Frequency-dependent dielectric measurements are performed using an LCR meter (NF Corporation ZM2376) for a frequency range of 1 kHz–1 MHz at room temperature. The AFM and PFM measurements are performed on a commercial MFP-3D 5 Asylum Research and Oxford Instruments Company AFM system (Asylum Research, Santa Barbara, CA, USA) using Dual AC Resonance Tracking (DART) PFM mode.

## RESULTS AND DISCUSSION

### STRUCTURE AND MORPHOLOGY

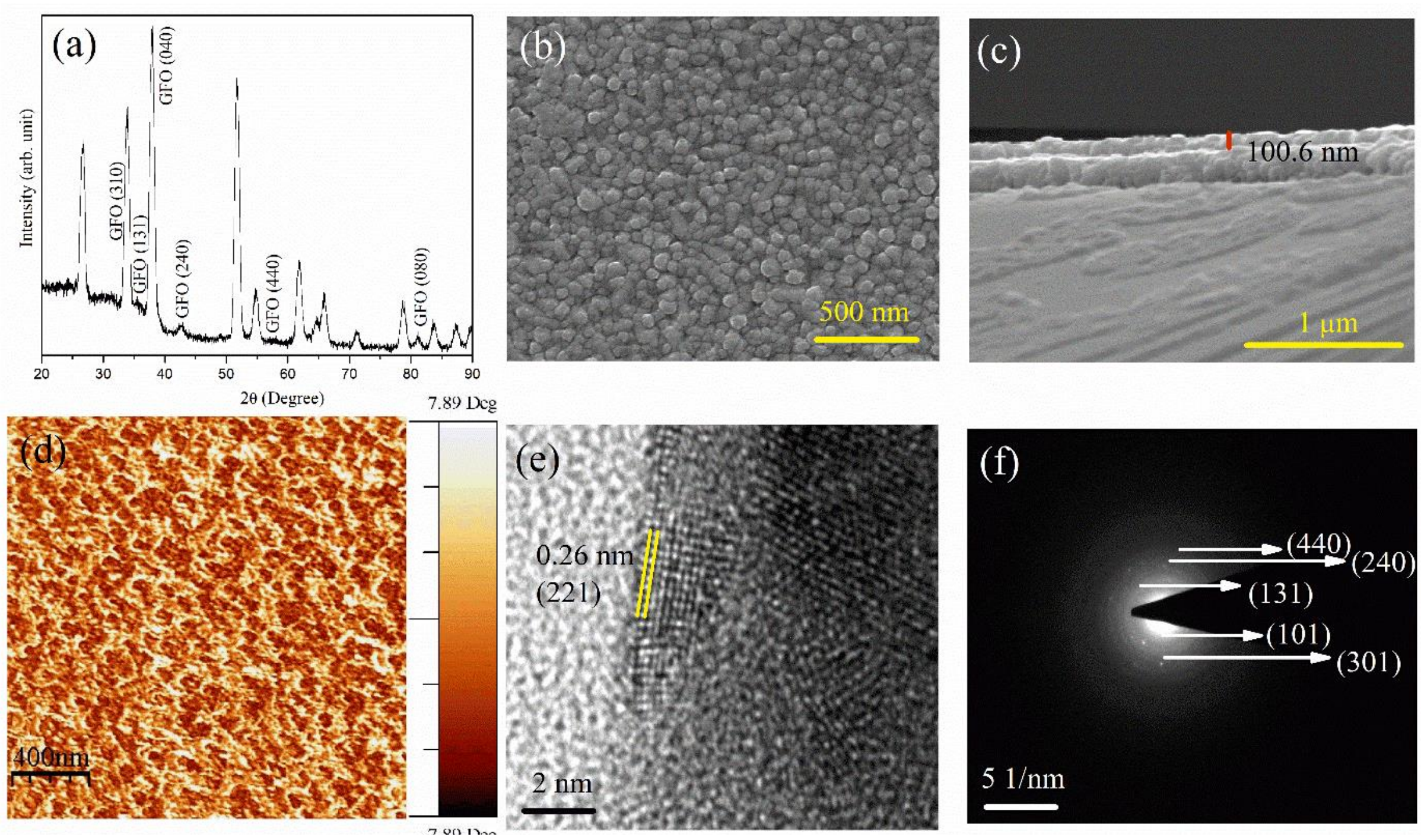

Figure 1: (a) The room temperature XRD data, (b) FE-SEM images, (c) Cross-sectional SEM image, (d) AFM image of GFO film of GFO films deposited on FTO substrate, (e) shows HRTEM, (f) depicts SAED data of GFO thin film

Fig. 1(a) shows the GI-XRD pattern of the GFO thin film deposited on a transparent conductive oxide FTO substrate, also acting as the bottom electrode. XRD peaks are indexed using JCPDS card no. 76-1005, confirming the formation of a polycrystalline GFO thin film with orthorhombic *Pc21n* symmetry. FESEM micrograph of the GFO thin films on FTO is shown in Fig. 1(b). The micrograph reveals a largely uniform deposition of GFO. Cross-sectional SEM demonstrates an average film thickness of ~100.6 nm, as shown in Fig. 1(c). Fig. 1(d) presents the surface topography analysis using atomic force microscopy, revealing a root-mean-square (RMS) roughness of 11.45 nm, suggesting uniform surface topography of the film.

The TEM micrographs with HRTEM and SAED patterns are presented in Fig. 1(e-f), respectively. The HRTEM (Fig. 1(e)) image reveals the lattice fringes of GFO (221) with a d-spacing of 0.26 nm, which is consistent with the reported literature [12]. The SAED pattern (Fig. 1(f)) also confirms the polycrystalline nature of GFO. The fringes corresponding to the plane of GFO are also confirmed by XRD.

The surface chemical states and elemental composition of the GFO film deposited on FTO were investigated using X-ray Photoelectron Spectroscopy (XPS). The survey spectrum (Fig. 2(a)) confirms the presence of Ga, Fe, and O peaks, explicitly identifying the constituent elements of the sample. Notably, the detection of Ga LMM and Fe LMM Auger peaks further confirms a successful deposition of the gallium ferrite phase, consistent with established literature[13]. High-resolution core-level spectra of Ga 2p (Fig. 2(b)) exhibit two prominent peaks at 1143.6 eV and 1116.3 eV, attributed to Ga $2p_{1/2}$ and Ga $2p_{3/2}$ spin-orbit coupling. The energy difference Δ = 27.3 eV matches the spin-orbit splitting characteristic of $Ga^{3+}$ ions, confirming the trivalent oxidation state of Ga in the film[14].

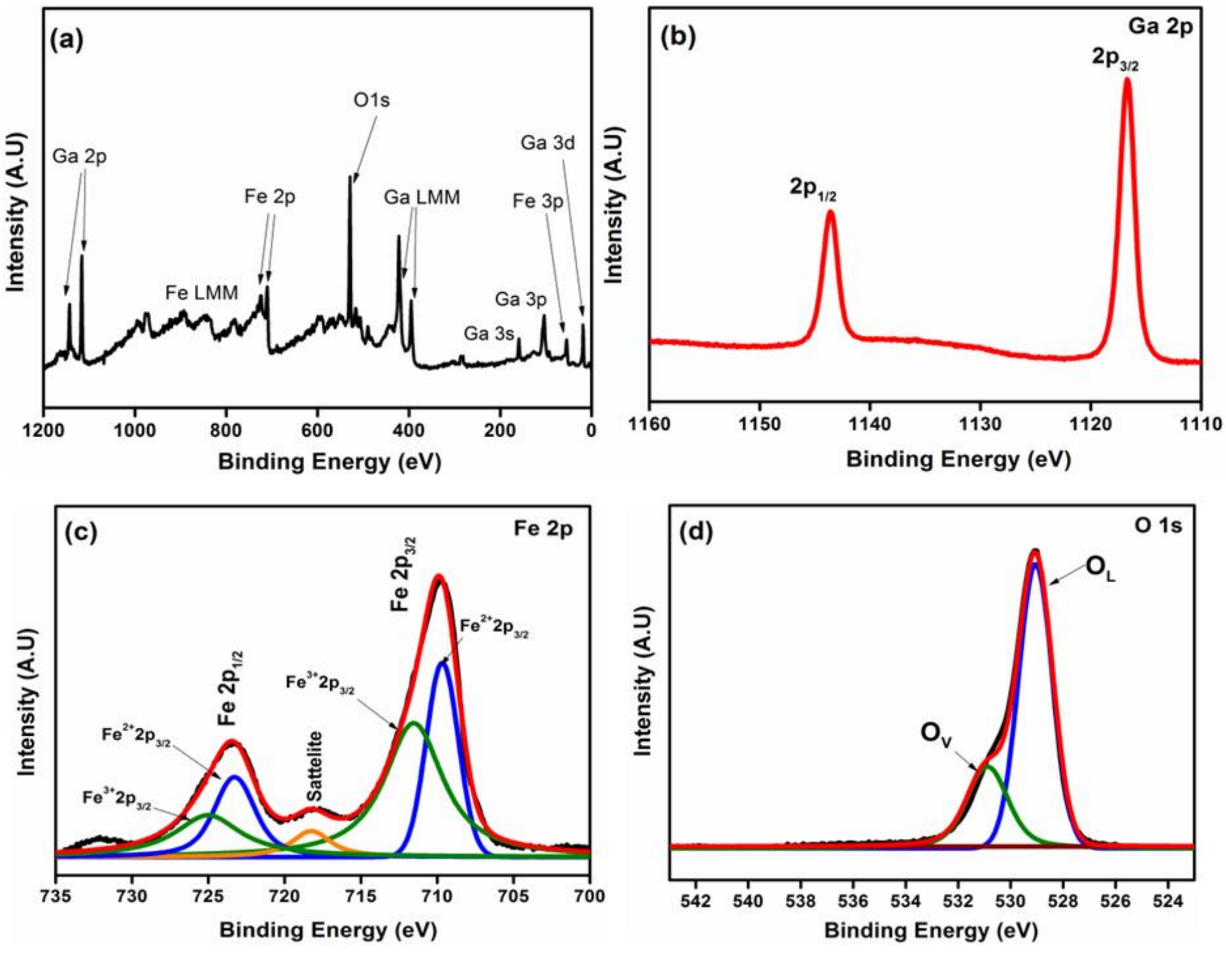

Figure 2: (a) Surface XPS scan of GaFeO3 thin film and the core level XPS profile of (b) Ga 2p, (c) Fe 2p and (d) O 1s

The Fe 2p core-level spectra (Fig. 2(c)) reveal a doublet at 723.1 eV attributed to Fe 2p1/2 and 709.3 eV, corresponding to Fe $2p_{3/2}$, accompanied by a distinct satellite peak at approximately 718.2 eV, which signifies a dominance of $Fe^{3+}$ species[15,16]. Further deconvolution of these peaks based on the contributions of $Fe^{3+}$ and $Fe^{2+}$ yields components centred around 709.1 eV and 722.4 eV for $Fe^{2+}$, and 711.2 eV and 725.3 eV for $Fe^{3+}$ [17]. Quantitative analysis of the area under the curves indicates an $Fe^{3+}/Fe^{2+}$ ratio of approximately 1.65, suggesting the film is comprised of roughly 63% $Fe^{3+}$ and 37% $Fe^{2+}$. This chemical environment suppresses leakage currents, which is conducive to enhanced magnetodielectric properties[18].

The O 1s core-level spectra (Fig. 2(d)) exhibit a strong lattice oxygen ($O_L$) peak at 529.5 eV, attributed to the Ga-O and Fe-O bonds within the crystal structure. An additional peak at 531.1 eV is assigned to oxygen vacancies ($O_V$)[19,20]. The calculated $O_L/O_V$ ratio of 2.67 indicates a low concentration of oxygen vacancies in the film. This stoichiometric quality is highly beneficial, as it limits the formation of $Fe^{2+}$, a correlation supported by the Fe spectrum in Fig. 2(c), thereby playing a vital role in suppressing leakage current and enhancing the multiferroic performance of the material[20].

## ELECTRICAL PROPERTIES

The microscopic ferroelectric and piezoelectric properties of the thin film are unambiguously confirmed using the PFM measurements. PFM provides the possibility of direct investigation of domain evolution in ferroelectrics under an external electric field[21]. When a bias voltage is applied between the conductive probe tip and bottom electrode, a localized electric field of several thousand kV/cm is generated, often exceeding the coercive field of the material. This allows PFM not only to image domain structures but also to induce local polarization switching within individual grains. Such switching minimizes the masking effect of leaky grain boundaries, which otherwise suppresses the intrinsic ferroelectricity[10,22]. To demonstrate the polarization reversal, a sequential positive and negative polling voltage of 6V is applied over a square region of 1μm × 1μm, during which the PFM phase and amplitude responses were simultaneously recorded. Fig. 3(a-c) shows out-of-plane piezoresponse at RT after the above-mentioned polling treatment. The phase images clearly show distinct contrast among the positive, negative and unpoled regions. Therefore, polarization can be reversed by switching the external electric field, and hence, the GFO polycrystalline film is ferroelectric at RT[23,24].

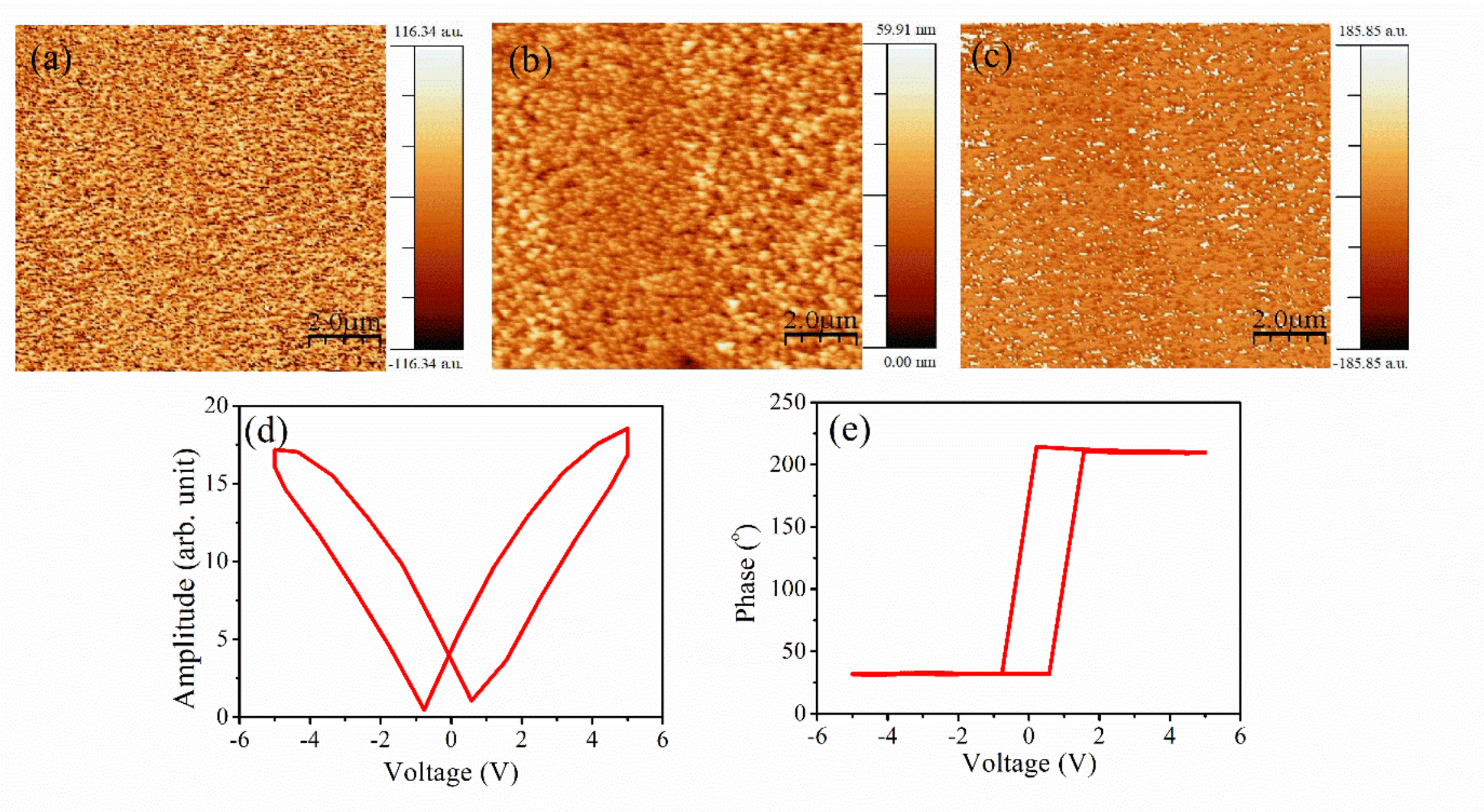

Figure 3: The piezoresponse force microscopy phase micrograph for GFO thin film at a bias field of (a) -6 V, (b) 6V, and (c) +6V, (d) amplitude and (e) phase hysteresis loops.

Furthermore, to confirm ferroelectric switching, the butterfly amplitude loop and phase as a function of DC voltage are shown in Fig. 3(d) and 3(e), respectively. The butterfly loop reveals the first harmonic signal under an applied DC bias field and is the signature of the piezoelectric response of the thin films. The piezoresponse tends to saturate at relatively high voltages, suggesting that the response is piezoelectric instead of electrostatic. The phase of the piezoresponse reverses with voltage. This reversal occurs beyond a coercive voltage, ~ 0.31 V on the negative side and ~1.07 V on the positive side, while the phase contrast is ~180°, clearly suggesting polarization switching and thus the ferroelectric character of our GFO thin films.

## DIELECTRIC PROPERTIES

Fig. 4(a) shows the frequency dependence of the real relative permittivity (ε′) of the thin film, while the inset presents the dielectric loss tangent (tan(δ)) at room temperature over 1 kHz–1 MHz. The decrease in permittivity with increasing frequency indicates dielectric dispersion. At low frequencies, interfacial and space-charge polarization arise from charge accumulation at interfaces and electrically heterogeneous regions. At higher frequencies, these slower polarization mechanisms cannot follow the alternating electric field. Notably, the relative permittivity remains remarkably high, approximately 1.7 k at 1 MHz, while the dielectric loss tangent is approximately 2.83, indicating high apparent permittivity accompanied by substantial energy dissipation. The dielectric response is influenced by relaxation processes, charge hopping, and electrical conduction. XPS analysis also confirms the coexistence of $Fe^{2+}$ and $Fe^{3+}$ states, which can facilitate electron hopping.

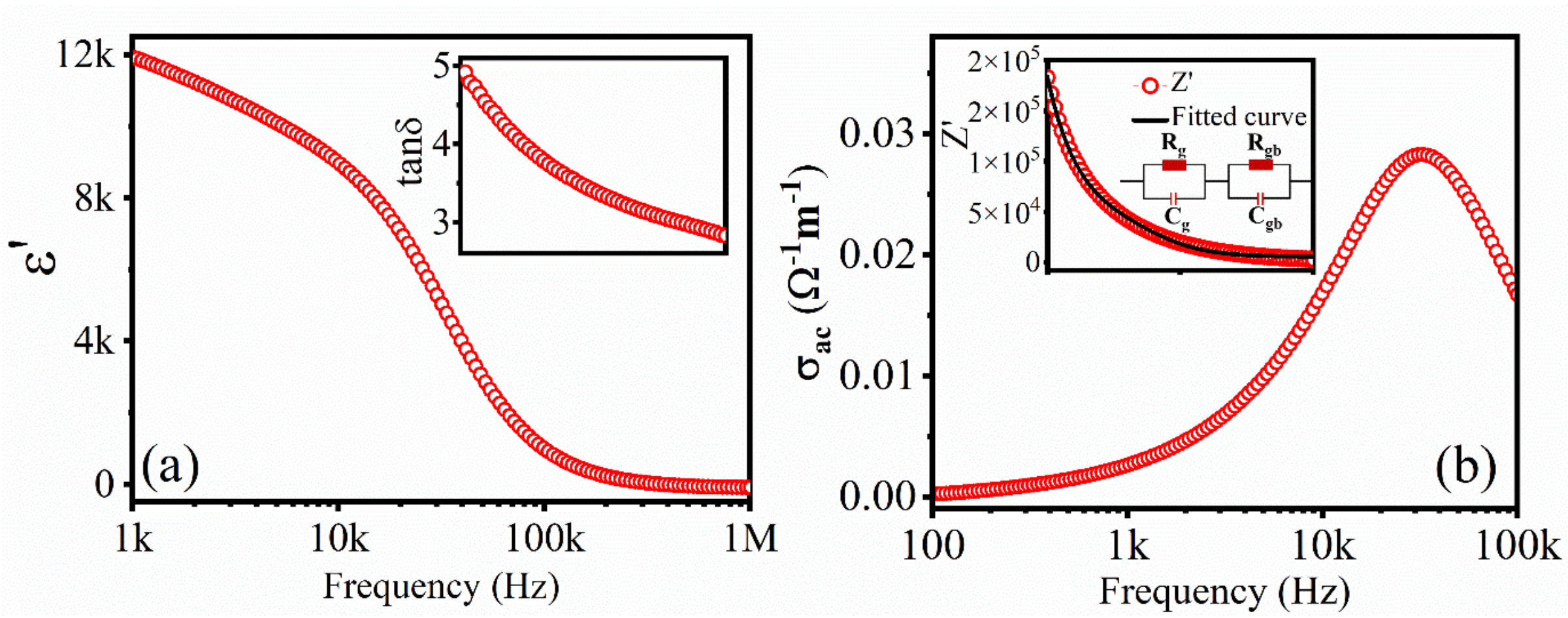

Figure 4: The frequency dependence of (a) the real part of complex permittivity, inset shows the tangent loss, (b) total conductivity at RT fitted curve of real part (Z') of complex impedance with frequency is shown in (b).

The frequency-dependent conductivity in Fig. 4(b) and the impedance spectra in its inset provide further insight into charge transport. The impedance spectra were fitted using an equivalent-circuit model[25,26], as shown in the inset of Fig. 4(b). The electrical conductivity is approximately $2.38 \times 10^{-4} \Omega^{-1} m^{-1}$ at 100 Hz, suggesting limited charge transport. Such behaviour is desirable for thin-film dielectric devices, as low conductivity can help suppress leakage currents. The fitted grain and grain-boundary resistances are approximately 0.04 MΩ and 2.49 MΩ, respectively. The quite high value of grain-boundary resistance indicates that grain boundaries act as barriers to charge transport and influence dielectric relaxation[4,27]. Overall, these findings highlight the potential of gallium ferrite thin films for integrated multifunctional devices. Their high apparent permittivity and relatively low conductivity are promising for practical applications, although further optimization is needed to reduce dielectric loss and energy dissipation.

## MAGNETIC AND MAGNETODIELECTRIC PROPERTIES

The room-temperature M–H loop, after subtracting the substrate contribution [Fig. 5(a)], exhibits a nearly saturated hysteresis response, confirming long-range magnetic ordering in the polycrystalline GFO thin film. This behaviour is consistent with the established ferrimagnetic structure of GFO, where interactions between Fe-containing sublattices generate a net magnetic moment, in agreement with previous reports on GFO and GFO-based thin films[28,29].

Following the confirmation of room-temperature polarization switching, the magnetodielectric (MD) response was investigated to assess the coupling between the magnetic and dielectric properties of the film, which is important for potential room-temperature multiferroic memory applications. The magnetodielectric factor can be derived from the equation[30,31],

$$\mathrm{MD}\ (\%) = \frac{\varepsilon'(H) - \varepsilon'(0)}{\varepsilon'(0)} \times 100\%$$

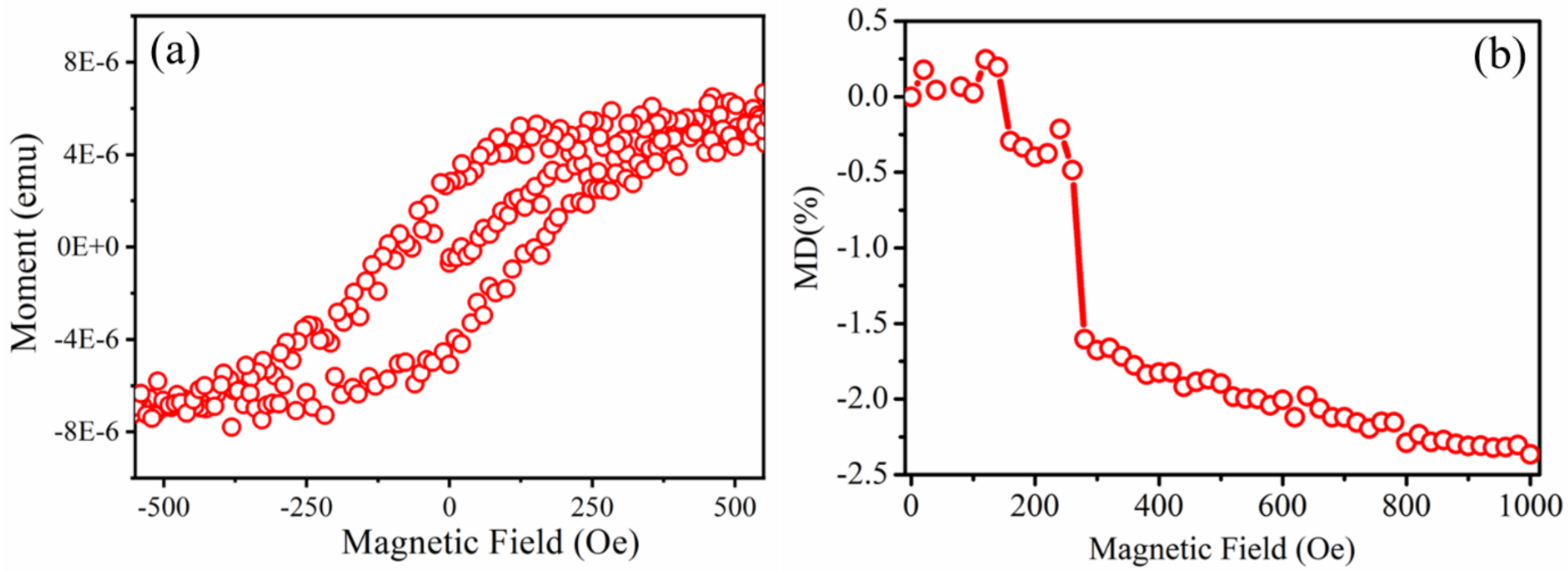

Figure 5: (a) the M-H hysteresis loop of GFO thin film at room temperature, (b) the variation of the magnetodielectric constant (MD) with changing applied magnetic field.

The dielectric constant is measured under magnetic fields ranging from 0 to 1 kOe, and the field-dependent MD factor is evaluated at 100 kHz [Fig. 5(b)]. Remarkably, an MD coefficient of 2.3% is obtained at a low magnetic field of only 1 kOe, representing a significant enhancement over previously reported values for other GFO morphologies. The MD response increases sharply in the low-field region at around 300 Oe[32]. The initial sharp rise may be associated with domain-wall displacement within the Fe-containing sublattices. In addition, substrate-induced lattice mismatch and structural constraints may contribute to the enhanced magnetodielectric response[7,33]. These results highlight the potential of GFO thin films for low-field multifunctional electronic devices.

## CONCLUSION

Polycrystalline $GaFeO_3$ thin films were fabricated on FTO substrates by pulsed laser deposition with an average thickness of ∼100.6 nm and investigated through complementary structural, chemical, electrical, dielectric, and magnetic measurements. XRD and electron microscopy confirm the formation of orthorhombic polycrystalline GFO without detectable crystalline secondary phases. XPS reveals $Ga^{3+}$ and mixed $Fe^{2+}/Fe^{3+}$ states. Out-of-plane PFM demonstrates reversible local switching with an approximately 180° phase contrast and coercive voltages of −0.31 V and +1.07 V, while room-temperature magnetic measurements reveal ferrimagnetic ordering. The film further exhibits a very high $\varepsilon' \approx 1700$ with a low $\tan\delta \approx 2.83$ at 1 MHz, and a strong magnetodielectric response of approximately 2.3% at 100 kHz under a 0.1 kOe magnetic field. The coexistence of room-temperature magnetic ordering, local polarization switching, and field-dependent dielectric response demonstrates the multifunctional character of the GFO thin film.

## ACKNOWLEDGMENT

We acknowledge financial support from S. N. Bose National Centre for Basic Sciences (SNBNCBS). We also want to gratefully acknowledge Mr. Deep Singha Roy, SRF, IACS, for his help with AFM and PFM analysis.